\documentclass[prl,letterpaper,twocolumn,showpacs]{revtex4}
\usepackage{times,xspace}
\usepackage{amsbsy,amssymb,amsmath,bm}
\usepackage{graphicx,color,epsfig,rotate}
\usepackage{fancyhdr}

\def\bbbc{{\mathchoice {\setbox0=\hbox{$\displaystyle\rm C$}\hbox{\hbox
to0pt{\kern0.4\wd0\vrule height0.9\ht0\hss}\box0}}
{\setbox0=\hbox{$\textstyle\rm C$}\hbox{\hbox
to0pt{\kern0.4\wd0\vrule height0.9\ht0\hss}\box0}}
{\setbox0=\hbox{$\scriptstyle\rm C$}\hbox{\hbox
to0pt{\kern0.4\wd0\vrule height0.9\ht0\hss}\box0}}
{\setbox0=\hbox{$\scriptscriptstyle\rm C$}\hbox{\hbox
to0pt{\kern0.4\wd0\vrule height0.9\ht0\hss}\box0}}}}

\newcommand{\ignore}[1]{}
\newcommand{\mComment}[1]{}
\newcommand{\gComment}[1]{}
\newcommand{\jComment}[1]{}
\newcommand{\rComment}[1]{}
\newcommand{\lComment}[1]{}

\renewcommand{\mComment}[1]{\textcolor{blue}{Manny: #1}}
\renewcommand{\gComment}[1]{\textcolor{red}{Gerardo: #1}}
\renewcommand{\jComment}[1]{\textcolor{green}{Jim: #1}}
\renewcommand{\rComment}[1]{\textcolor{magenta}{Ray: #1}}
\renewcommand{\lComment}[1]{\textcolor{purple}{Rolando: #1}}

\begin{document}

\title{Geometric Frustration and Dimensional Reduction at a Quantum Critical
Point}
\author{C. D. Batista$^{1}$, J. Schmalian$^{3}$, N. Kawashima$^{4}$, P.
Sengupta$^{1,2}$, S. E. Sebastian$^{5,6}$, N. Harrison$^{2}$, M. Jaime$^{2}$
and I. R. Fisher$^{5}$}
\affiliation{$^{1}$Theoretical Division and $^{2}$MPA-NHMFL, Los Alamos National
Laboratory, Los Alamos, NM 87545 \\
$^3$ Department of Physics and Astronomy, Iowa State University and Ames
Laboatory, Ames IA50011 \\
$^4$ Institute for Solid State Physics, University of Tokyo, Kashiwa, Chiba
227-8581, Japan \\
$^5$ Geballe Laboratory for Advanced Materials and Department of Applied
Physics, Stanford University, Stanford, CA 94305 \\
$^{6}$ Cavendish Laboratory, University of Cambridge, Madingley Road,
Cambridge CB3 0HE, UK }
\date{\today }

\begin{abstract}
We show that the spatial dimensionality of the quantum critical point
associated with Bose--Einstein condensation at $T=0$ is reduced when the
underlying lattice comprises a set of layers coupled by a frustrating
interaction. Our theoretical predictions for the critical temperature as a
function of the chemical potential correspond very well with recent
measurements in BaCuSi$_{2}$O$_{6}$ [S. E. Sebastian \textit{et al}, Nature 
\textbf{411}, 617 (2006)].
\end{abstract}

\maketitle

The universality class of a critical point (CP) depends on a few properties such
as the symmetry of the underlying model, the range of the interactions, the
number of components of the order parameter (OP), and the space
dimensionality $d$\cite{Stanley}. It is well known that for highly
anisotropic systems such as weakly coupled layers, the universality class
changes when the system approaches the CP. A dimensional
crossover takes place: the effective dimensionality is reduced {\it beyond} a
certain distance from the CP, determined by the weak
inter--layer interaction. Sufficiently close to the CP the
transition is however three--dimensional. In contrast to this common
behavior, the dimensional reduction (DR) discussed in this paper occurs when
the system \emph{approaches} a gaussian quantum critical point (QCP). The
source of this qualitative difference is in the nature of the interlayer
coupling. We show that the interlayer coupling vanishes right at the
QCP for a chemical potential tuned Bose--Einstein condensation
(BEC) of interacting bosons. We then argue that this effect is relevant for
the field tuned QCP of a \textit{geometrically frustrated} quantum magnet.

Although geometric frustration has previously been invoked \cite{Stockert98}
as a mechanism for DR, zero--point fluctuations are expected to restore the
inter-layer coupling \cite{Shender82}, as explicitly shown by Maltseva and
Coleman \cite{Maltseva05}. In distinction to this expectation, we show that
this coupling is suppressed near the BEC--QCP, relevant to spin dimer
systems in a magnetic field. In this case, the spatial dimensionality of the
gaussian QCP is $d=2$. Interactions between either thermally excited or
quantum condensed bosons induce a crossover to $d=3$ away from the QCP. Key
to this result is the observation that zero--point phase fluctuations of the
OP are suppressed near a chemical potential tuned BEC. The first
experimental evidence of this phenomenon was found very recently by
measuring critical exponents of a field induced QCP in BaCuSi$_{2}$O$_{6}$ 
\cite{Suchitra06}.

We start by presenting the rigorous result for the model of a chemical potential
tuned BEC where the bosons are located on the sites of a body--centered
tetragonal (bct) lattice. In the second part of the paper we discuss the relevance
of this results for $S=$ $\frac{1}{2}$ spins forming dimers on a bct
lattice, closely approximating the case of BaCuSi$_{2}$O$_{6}$ \cite 
{Sparta04,Samulon06}. This offers a quantitative explanation of the observed
DR in this system\cite{Suchitra06}.

We start from the Hamiltonian of interacting bosons 
\begin{equation}
H_B=\sum_{\mathbf{k}}\left( \varepsilon _{\mathbf{k}}-\mu \right) a_{\mathbf{ 
k} }^{\dagger }a_{\mathbf{k}}+u \sum_{i}n_{i}n_{i}  \label{Hb}
\end{equation}
where $n_{i}=a_{i}^{\dagger }a_{i}$ is the local number operator of the
bosons and $a_{\mathbf{k}}^{\dagger }=\sum_{i}a_{i}^{\dagger }e^{i\mathbf{\
k\cdot R}_{i}}/\sqrt{N}$ the corresponding creation operator in momentum
space. The tight binding dispersion for nearest neighbor boson hopping on
the bcc lattice is 
\begin{equation}
\varepsilon _{\mathbf{k}}=\varepsilon _{\parallel }\left( \mathbf{k}
_{\parallel }\right) +2t_{\perp }\gamma \left( \mathbf{k}_{\parallel
}\right) \cos k_{z}  \label{tb2}
\end{equation}
where $\mathbf{k}_{\parallel }=\left( k_{x},k_{y}\right) $ refers to the in
plane momentum. $\varepsilon _{\parallel }\left( \mathbf{k}_{\parallel
}\right) =t_{\parallel }\left( 2+\cos k_{x}+\cos k_{y}\right) $ is the
in-plane dispersion while $\gamma \left( \mathbf{k}_{\parallel }\right) =
\cos \frac{k_{x}}{2}\cos \frac{k_{y}}{2}$. For $t_{\parallel }$, $t_{\perp
}>0$ and $t_{\parallel}>t_{\perp }/2$, a BEC takes place at $\mathbf{K} 
_{\parallel }=\left( \pi ,\pi \right)$. Since $\gamma \left( 
\mathbf{K}_{\parallel }\right) =0$, the dispersion $\varepsilon _{\mathbf{k}}
$ at the condensation momentum is independent of $k_{z}$. In case of the
ideal Bose gas ($u=0$) this implies for $T=0$ that different layers decouple
completely. Only excitations at finite $T$ with in-plane momentum away from
the condensation point can propagate in the $z$-direction. This behavior
changes as soon as one includes boson-boson interactions ($u>0$). States in
the Bose condensate scatter and create virtual excitations above the
condensate that are allowed to propagate in the $z$-direction. These
excitations couple to condensate states in other layers\cite{Maltseva05}.
The condensed state of interacting bosons is then truly three dimensional,
even at $T=0$.

The above argument for "dimensional restoration" due to interactions does
not apply in case of chemical potential tuned BEC. In this case, the number
of bosons at $T=0$ is strictly zero for $\mu <0$, i.e. before BEC sets in.
The absence of particles makes their interaction mute and one can approach
the QCP arbitrarily closely without coherently coupling different layers.
While the Bose condensed state for $\mu >0$ and the entire regime for $T>0$
is three dimensional, the decoupling for $\left( \mu <0,T=0\right) $ has
dramatic consequences. We show that the BEC transition temperature varies as 
\begin{equation}
T_{c}\propto \mu \ln \left( \frac{t_{\parallel }}{\mu }\right) /\ln \ln 
\frac{t_{\parallel }}{\mu }.  \label{Tc}
\end{equation} 
$T_{c}$ $\propto \mu^{2/d}$ holds instead for an isotropic Bose system in $d>2$. 
Despite the fact that different layers are coupled at finite $T$
the BEC-transition temperature,  Eq.(\ref{Tc}), depends on $\mu $ just like
the Berezinskii-Kosterlitz-Thouless (BKT) transition temperature of a two
dimensional system\cite{FisherHohenberg88}.

The renormalization group (RG) calculation used to obtain this result (a one-loop RG 
calculation in analogy to Ref.\cite{FisherHohenberg88})
shows that the finite temperature transition is a classical 
$3$-$d$ $XY$ transition, not a BKT transition. We
conclude, therefore, that the $T=0$ QCP of chemical potential tuned BEC
with three dimensional dispersion, Eq.(\ref{tb2}), is strictly two
dimensional. The system then crosses over to be three dimensional for $\mu >0
$ or $T>0$, where the density of bosons becomes finite and boson-boson
interactions drive the crossover to $d=3$. The transition temperature of
this three dimensional BEC is given by the two-dimensional
result, Eq.(\ref{Tc}). It is important to stress
that the vanishing density for $\left( \mu <0,T=0\right) $ implies that
these results are not limited to weakly interacting bosons\cite{Sachdev}.

\begin{figure}[tbh]
\vspace*{-0.4cm} 
\hspace*{0cm}  
\includegraphics[angle=0,width=9cm]{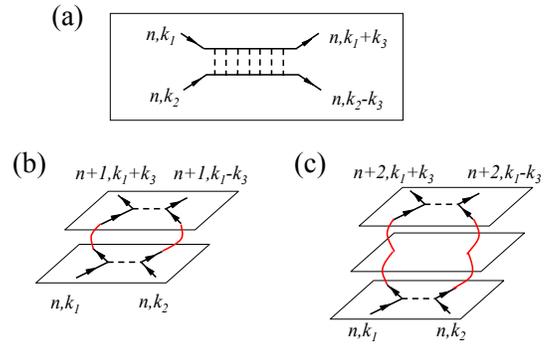} 
\vspace{-2.0cm}
\caption{(Color online)(a) Ladder diagrams that provide the dominant contribution to the
intra--layer scattering in the low density regime \cite{Beliaev}.
(b) and (c) leading order diagrams that contribute to the coherent
inter--layer hoppings $t_{\perp ,1}^{\ast }$ and $t_{\perp ,2}^{\ast }$.}
\label{diagrams}
\end{figure}

The detailed derivation of Eq.(\ref{Tc}) using the RG approach 
will be presented in a separate publication \cite{Batista07}. Here we present 
a heuristic derivation of the same result 
based on an approach introduced by Popov 
\cite{Popov83} and further explored by Fisher and Hohenberg \cite 
{FisherHohenberg88}: infrared divergencies are cut-off for momenta $ 
k<k_{0}\simeq \sqrt{\mu /t_{\parallel }}$. We analyze the interacting Bose
system in the disordered phase and perform an expansion in the interlayer
hopping amplitude $t_{\perp }/t_{\parallel }$. Dominant interactions at low
density are given by ladder diagrams (see Fig.\ref{diagrams}a), yielding a
renormalized boson interaction (i.e. the scattering matrix for bosons in the
same layer) \cite{Beliaev}: 
\begin{equation}
v_{0}^{-1}=\frac{1}{4}\int_{k_{0}}\frac{d^{2}k_{\parallel }}{4\pi
^{2}\varepsilon _{\parallel }\left( \mathbf{k}_{\parallel }\right) }\propto 
\frac{\ln k_{0}^{-1}}{t_{\parallel }},
\end{equation} 
for $u\rightarrow \infty $ (hard core bosons). The bare interlayer coupling 
leads to scattering of bosons between different layers. The corresponding
scattering matrices between neighboring layers, $v_{1}$ (see Fig.\ref 
{diagrams}b), and second neighbor layers, $v_{2}$ (see Fig.\ref{diagrams}c)
are then given as ($l=1,2$) 
\begin{equation}
v_{l}\simeq -\left( \frac{t_{\perp }}{t_{\parallel }}\right) ^{2l}\frac{ 
t_{\parallel }}{\ln k_{0}^{-1}},
\end{equation} 
where the overall negative sign results from the fact that the lowest order
contribution to $v_{1,2}$ are of order $v_{0}^{2}$. This interlayer coupling
is on the interaction level, and leads to new non-local interaction terms $ 
v_{l}n_{i}n_{i+l\mathbf{e}_{z}}$ in the low energy Hamiltonian of the model.
The origin of these couplings are $T=0$ quantum fluctuations of the
interacting Bose system. Pairs of boson propagate as virtual excitations
between layers and mediate the non-local boson-boson coupling\cite 
{Maltseva05}. It is crucial to observe that, no coherent boson hopping $ 
t_{\perp ,l}^{\ast }a_{\mathbf{k}_{\parallel },n}^{\dagger }a_{\mathbf{k} 
_{\parallel },n+l}$ between layers emerges for $T=0$. 

$H_B$ is invariant with respect to the discrete $Z_{2}$--symmetry: $ 
k_{x}\rightarrow -k_{x}+ 2\pi$ and $k_{z}\rightarrow k_{z}+\pi $. As long as
this symmetry is intact, no term $t_{\perp ,1}^{\ast }\cos \left(k_{z}\right)
$ in the dispersion is allowed, while coherent hopping between second
neighbor layers with $t_{\perp ,2}^{\ast }\cos \left( 2k_{z}\right) $ does
not break the $Z_{2}$-symmetry. To determine these coherent interlayer
hoppings $t_{\perp ,l}^{\ast }$ we perform a mean field (MF) theory of the
low energy problem with interlayer interactions $v_{l}$. We approximate $ 
v_{l}n_{i}n_{i+l\mathbf{e}_{z}}\rightarrow v_{l}\left\langle a_{i}^{\dagger
}a_{i+l\mathbf{e}_{z}}\right\rangle a_{i}^{\dagger }a_{i+l\mathbf{e}_{z}}$
and obtain 
\begin{equation}
t_{\perp ,l}^{\ast }=v_{l}\int \frac{d^{2}k_{\parallel }}{4\pi ^{2}}
\left\langle a_{\mathbf{k}_{\parallel },n}^{\dagger }a_{\mathbf{k}
_{\parallel },n+l}\right\rangle .  \label{t}
\end{equation}
The expectation values $\left\langle a_{\mathbf{k}_{\parallel },n}^{\dagger
}a_{\mathbf{k}_{\parallel },n+l}\right\rangle $ for the single particle
overlap between neighboring layers are determined self-consistently. As
expected, we find $t_{\perp ,2}^{\ast }\left( T=0\right) =0$ and $t_{\perp
,1}^{\ast }\left( T\right) =0$. The former result reflects the fact that no
coherent motion is possible at $T=0$, while the latter is caused by the $ 
Z_{2}$-symmetry, forcing the hopping between nearest neighbor layers to
vanish at all $T$. The solution of Eq.(\ref{t}) for the coherent second
neighbor hopping is 
\begin{equation}
t_{\perp ,2}^{\ast }\simeq v_{2}\left( \frac{t_{\perp }}{t_{\parallel }}
\right) ^{2}\frac{T}{t_{\parallel }}\ln \frac{T}{t_{\parallel }k_{0}^{2}}.
\end{equation}
Using the above result for $v_{2}$ it then follows $t_{\perp ,2}^{\ast
}\simeq \left( \frac{t_{\perp }}{t_{\parallel }}\right) ^{6}\frac{T\ln
T/\mu }{\ln t_{\parallel }/\mu }$. Since the density of bosons is $\rho
\simeq T\ln \left( T/\mu \right) /t_{\parallel }$, one sees that thermally
excited bosons induce a coherent hopping between second neighbor layers,
i.e. $t_{\perp ,2}^{\ast }$ $\propto \rho /\ln \left( t_{\parallel }/\mu
\right) $. While the amplitude of this coherent hopping is small, the finite 
$T$ transition will be three dimensional. The ordering temperature of this $3
$-$d$ $XY$ transition is given by the MF condition: 
\begin{equation}
\mu_c = v_0 \rho,  \label{mumf}
\end{equation}
and as usual for strongly anisotropic systems, its value is given by the
characteristic temperature scale of the in plane ordering. Since $t_{\perp
,2}^{\ast } \ll \epsilon_{\parallel} (k_0)$, the $d=2$ fluctuations
dominate the magnitude of $T_c$ at very low densities resulting in
Eq.(\ref{Tc}). For the same reason, we also obtain $d=2$ expressions
for: 
\begin{eqnarray}
\rho(T=0,\mu) &\propto& \mu \ln{\frac{\mu}{t_{\parallel}}}  \notag \\
\rho(T,\mu=0) &\propto& \frac{T}{t_{\parallel}} \ln({\ln{\frac{ 
t_{\parallel}}{T}}})  \label{rho}
\end{eqnarray}

Based on these results we next address the origin of DR in the frustrated
magnet BaCuSi$_{2}$O$_{6}$ \cite{Suchitra06}. We start from a Heisenberg
Hamiltonian of $S=$ $\frac{1}{2}$ spins dimers on a bct lattice, 
closely approximating the case of BaCuSi$_{2}$O$_{6}$ 
\cite{Sparta04,Samulon06}. The dominant Heisenberg interaction, $J\sum_{ 
\mathbf{i}}\mathbf{S}_{i1}\cdot \mathbf{S}_{i2}$, is between spins on the
same dimer $i$. Since there are two low energy states in an applied magnetic
field, the singlet and the $S_{i1}^{z}+S_{i2}^{z}=1$ triplet, we can
describe the low energy sector with hard--core bosons. The triplet state
corresponds to an effective site $i$ occupied by a boson while the singlet
state is mapped into the empty site \cite{Giamarchi99,Jaime04}. The
resulting low energy effective Hamiltonian corresponds to a gas of
interacting (infinite on--site repulsion) canonical bosons, as given in Eq.( 
\ref{Hb}). The number of bosons (number of triplets) equals the
magnetization along the $z$--axis. The chemical potential $\mu =g\mu
_{B}(H-H_{c})$ is determined by the applied magnetic field $H$ and the
critical field $g\mu _{B}H_{c}=J-2J^{\prime}$. The hoppings $t_{\parallel
}=J^{\prime }$ and $t_{\perp }=J^{\perp}$ are determined by the inter--dimer
exchange interactions between spins on the same bilayer, $J^{\prime}\simeq 6$ 
K \cite{Jaime04,Suchitra05,Ruegg06} and on adjacent bilayers, $J^{\perp} <
J^{\prime}$. The modulation of the BaCuSi$_{2}$O$_{6}$ lattice structure along
the $c$--axis leads to an alternation of two non--equivalent bilayers
A and B, with intra--dimer interactions $J_A=49.5(1)$K and $J_B=54.8(1)$K 
\cite{Ruegg06,Samulon06}. This alternation reduces the  
magnitude of the residual non-frustrating inter--layer  
couplings characteristic to all real systems \cite{note0}, while the  
principal treatment of BaCuSi$_{2}$O$_{6}$ presented here remains unaffected.

\begin{figure}[!htb]
\vspace*{1.0cm} 
\hspace*{0cm} 
\includegraphics[angle=90,width=11cm]{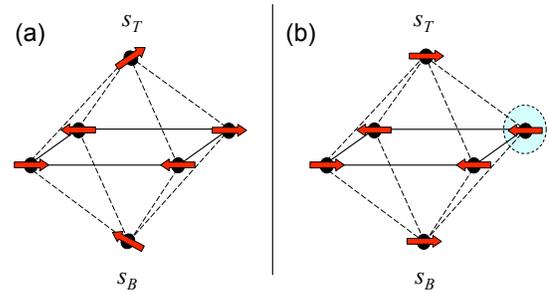} 
\vspace{-2.5cm}
\caption{(Color online)(a) The perfect antiferromagnetic (AF) order of the four spins in
the square plaquette precludes an effective coupling between $S_B$ and $S_T$ 
. (b) A phase fluctuation of the AF OP induces an effective ferromagnetic
coupling between $S_B$ and $S_T$. For BaCuSi$_{2}$O$_{6}$, each site
represents a dimer.}
\label{ofo}
\end{figure}

The correspondence between the quantum spin model for BaCuSi$_{2}$O$_{6}$
with the boson model of Eq.(\ref{Hb}) allows us to interpret $T_{c}$ of Eq.(\ref 
{Tc}) as the phase boundary as a function of $\mu =g \mu_{B}(H-H_{c})$. At
this phase transition, we also expect that the $Z_{2}$ symmetry will be
broken as well. It is interesting to analyze the dimensional crossover and
the coupling between second neighbor layers directly in the spin language.
For classical spins $\mathbf{S}_{\mathbf{i}}$ at $T=0$, the frustrated
nature of $J^{\perp }$ produces a perfect decoupling of the OP's ($XY$
staggered magnetization) on different layers. However, this decoupling is
unstable with respect to quantum or thermal fluctuations\cite{Maltseva05}.
Either of these fluctuations induces an effective inter--layer coupling via
an order from disorder mechanism as illustrated in Fig.\ref{ofo}. When the
sum of the four spins on a given plaquette, $\mathbf{S}_{P}$, is exactly
equal to zero the coupling between that plaquette and the spins which are
above ($\mathbf{S}_{T}$) and below ($\mathbf{S}_{B}$) cancels out. However,
the effect of phase fluctuations is to produce a net total spin on the
plaquette, $\mathbf{S}_{P}\neq 0$. Since $\mathbf{S}_{P}$ is
antiferromagnetically coupled with $\mathbf{S}_{T}$ and $\mathbf{S}_{B}$
(see Fig.\ref{ofo}b), an effective ferromagnetic (FM) interaction results
between $\mathbf{S}_{T}$ and $\mathbf{S}_{B}$, i.e. between second neighbor
layers. 

The defining characteristic of the BEC quantum phase transition is that it
is driven by amplitude fluctuations of the OP, in contrast to the $XY$-like
transition that is driven by phase fluctuations. This difference is vital to
the effective coupling at the QCP: it vanishes due to its quadratic
dependence on the amplitude of the OP. The remarkable consequence is a DR of
the gaussian QCP from $d=3$ to $d=2$.

\begin{figure}[!htb]
\vspace*{-1.0cm} 
\hspace*{0.0cm}  
\includegraphics[angle=0,height=9cm,width=8cm]{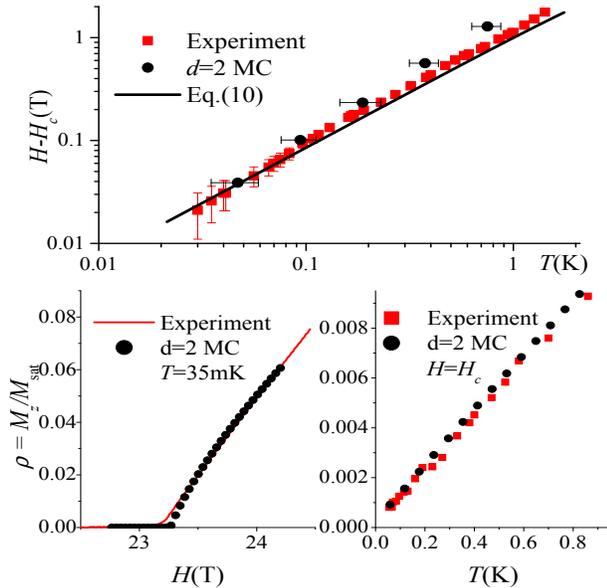} \vspace{-0.8cm}
\caption{(Color online) (a) Phase boundary near the QCP measured in BaCuSi$_2$O$_6$ 
\protect\cite{Suchitra06} compared to the curves obtained from a MC
simulation and Eq.(\protect\ref{muc}) for a $d=2$ gas of hard--core bosons
on a square lattice with $t_{\parallel}=J^{\prime}=6$K.(b) Similar
comparison for $\protect\rho(T=30\mathrm{mK},\protect\mu=H-H_c)$ and $ 
\protect\rho(T=30\mathrm{mK},H=H_c)$. We have neglected the density of
bosons on the B-bilayers because $J_2-g\protect\mu_B H \gg |t^*_{\perp,l}|$
as long as $H$ is not close to $J_2/g\protect\mu_B \simeq H_c + 3.4$T.}
\label{hcvstc}
\end{figure}

Our previous analysis shows that the non--universal prefactors of Eqs.(\ref{Tc}) 
and (\ref{rho}) can be determined to high accuracy by using a theory
for a strictly $d=2$ system. The RG and MF
approaches used to describe the quasi-condensate phase of a weakly
interacting two--dimensional Bose gas \cite{Popov83} give the proper
universality class and generic $T$ vs. $\mu $ dependence, yet are
quantitatively not adequate for realistic densities. The limitation of these
treatments arises from the insensitivity of the
size of the critical region $\Delta T$ (where the fluctuation corrections
associated with the BKT transition are important) to
the smallness of the interaction: $\Delta T/T_{c}\sim 1/\ln {t_{\parallel
}/v_{0}}$. This limitation is, however, overcome by our use of 
Monte Carlo (MC) simulations to obtain the non--universal constants that appear in 
Eq.(\ref{Tc}), and the results of Prokof'ev and Sustinov \cite{Prokofev02} 
who computed these constants explicitly for realistic low densities and weak interactions, 
obtaining the following expression
for the phase boundary of the quasi-condensate \cite{Prokofev02}: 
\begin{equation}
\mu _{c}=\frac{v_{0}T}{\pi J^{\prime }}\ln {\frac{J^{\prime }\xi _{\mu }}{ 
v_{0}}}  \label{muc}
\end{equation} 
where $\xi _{\mu }=13.2\pm 0.4$. In Fig.~\ref{hcvstc}a, we compare the
experimental data for BaCuSi$_{2}$O$_{6}$ \cite{Suchitra06} with the result
of Eq.(\ref{muc}) and Monte Carlo (MC) simulations of
hard core bosons on a square lattice ($L\times L$ with $L=32$) with hopping 
$t_{\parallel }=J^{\prime}=6$K \cite{Suchitra05,Ruegg06}. The agreement is good for 
$T\lesssim 200$mK ($\rho \lesssim 0.02$) but, as expected, there is a significant deviation at
higher temperatures (densities). The fact that the measured $T_{c}$ becomes
significantly higher than the MC result at higher temperatures indicates
that neglecting the effective inter-layer tunneling is no longer valid in
BaCuSi$_{2}$O$_{6}$ for $\rho \gtrsim 0.02$. Fig.~\ref{hcvstc}b shows a
similar comparison for $\rho (\mu ,T\simeq 0)$ and $\rho (\mu =0,T)$ [see
Eqs.(\ref{rho})]. Again, we compare the experimental data against the MC
simulation because the MF approximation that leads to Eqs.(\ref{rho}) is
adequate to determine the generic $\mu $ and $T$ dependence, but cannot reproduce
the non-universal constants. Our theory also predicts a linear
dependence of the specific heat $C(T,H_{c})$ and the nuclear relaxation time 
$1/T_{1}(T,H_{c})$ as a function of $T$ at the QCP of BaCuSi$_{2}$O$_{6}$.

To compute the exponent of the next order correction to Eq.(\ref{Tc}) we
note that the effective boson--boson interaction $v_0(\rho )$ is obtained as
an expansion in the small parameter $\rho ^{1/2}$ \cite{Beliaev}: ${\tilde v} 
_0(\rho )=v_{0}(1+ \alpha \rho ^{1/2}+...) $. While the first term in this
expansion is determined by the ladder diagrams of Fig.\ref{diagrams}a,
higher order diagrams contribute to the second term. The MF relation (\ref 
{mumf}) implies that the next order correction to Eq.(\ref{muc}) is
proportional to $T^{3/2}$.  The value of $u_1$ determines the crossover
 between the linear regime consistent with a $d=2$-QCP and the $T^{3/2}$
 regime characteristic of a $d=3$ BEC. Such a crossover was reported in BaCuSi 
 $_2$O$_6$ \cite{Suchitra06}. 
By following a similar procedure, we can demonstrate in general that the
phase boundary equation of a $d$-dimensional bosonic system that comprises $ 
d-1$-dimensional regions coupled via a frustrated interaction is $\mu_c
\simeq A T^{(d-1)/2} + B T^{d/2}$ for low enough $\rho$. 

Our mapping of the spin problem to the boson model $H_{B}$ is based on the
assumption that only the lowest triplet and the singlet modes are important
at low energies. The low density expansion for the boson problem is then
well justified, as the $XY$-symmetry of the original spins $\mathbf{S}_{i}$
is directly responsible for the charge conservation of $H_{B}$. Recently, it
was shown by R\"{o}sch and Vojta \cite{Rosch} that the inclusion of the two
higher triplet modes generates a small coherent second neighbor hopping of
low energy triplets between layers $t_{\perp ,2}^{\ast }\simeq J_{\perp
}^{6}/J^{5}$. This interesting effect restores the $d=3$ character of the
spin problem. For realistic values of $J=49.5(1)$K and $J_{\perp }<J^{\prime
}$, we find that $J_{\perp }^{6}/J^{5}<0.1$mK in BaCuSi$_{2}$O$_{6}$. This
implies that the mechanism discussed in our paper is still dominant for all
experimentally accessible temperatures $T\gtrsim 30$mK. Moreover, the U(1)-symmetry breaking
terms induced by dipolar interactions will produce a crossover to QCP with
discrete symmetry at $T\sim 10$mK \cite{Suchitra006} before the mechanism of
Ref.\cite{Rosch} sets in. Finally, the inevitable presence of finite
non-frustrated couplings in real systems  will eventually restore the three
dimensional behavior below some characteristic temperature $T_{0}$ (for
BaCuSi$_{2}$O$_{6}$ we estimate $T_{0}<30$mK\cite{note0}). We stress that
our theoretical results for $H_{B}$ [Eq.(\ref{Hb})] are not affected by
these considerations. There exists a non-trivial three dimensional
interacting many body system with a strictly $d=2$ QCP.

In summary, we demonstrate that the dimensionality of the BEC-QCP is $d = 2$ 
when the inter-layer coupling is frustrated. However, this coupling is relevant 
for changing the thermodynamic phase transition from BKT type to the 3d-XY  
universality class. These results explain quantitatively, and  
without free parameters, the DR manifested in the experimentally  
measured quantum critical exponents of BaCuSi$_{2}$O$_{6}$
\cite{Suchitra06}.

We thank A. J. Millis, N. Prokof'ev and M. Vojta for helpful discussions. LANL is
supported by US DOE under Contract No. W-7405-ENG-36. Ames Laboratory, is
supported by US DOE under Contract No. W-7405-Eng-82. S.E.S. and I.R.F are
supported by the NSF, DMR-0134613. Work performed at the NHMFL is supported by
the NSF (DMR90-16241), the DOE and the State of Florida. Part of the computational 
work wasexecuted on computers at the Supercomputer Center, ISSP,
University of Tokyo.


\begin{thebibliography}{99}
\bibitem{Stanley} H. E. Stanley,``\textit{Introduction to Phase Transitions
and Critical Phenomena},(Oxford University Press, Oxford, 1987).

\bibitem{Stockert98} O. Stockert \textit{et al}, Phys. Rev. Lett. \textbf{80}
, 5627 (1998); N. D. Mathur \textit{et al}, Nature (London) \textbf{394}, 39
(1998).

\bibitem{Shender82} E. Shender, Sov. Phys. JETP \textbf{56}, 178 (1982); C.
L. Henley, Phys. Rev. Lett. \textbf{62}, 2056 (1989).

\bibitem{Maltseva05} M. Maltseva and P. Coleman, Phys. Rev. B \textbf{72},
174415 (2005).

\bibitem{Suchitra06} S. E. Sebastian \textit{et al}, Nature \textbf{411},
617 (2006).

\bibitem{Sparta04} K. M. Sparta and G. Roth, Act. Crys. B. \textbf{60}, 491
(2004).

\bibitem{Samulon06} E. Samulon \textit{et al.}, Phys. Rev. B \textbf{73},
100407(R) (2006).

\bibitem{FisherHohenberg88} D. S. Fisher and P. C. Hohenberg, Phys. Rev. B 
\textbf{37}, 4936 (1988).

\bibitem{Sachdev} For a discussion of this issue see: S. Sachdev, Quantum
Phase Transitions (Cambridge Univ. Press 1999).

\bibitem{Batista07} J. Schmalian and C. D. Batista, unpublished.

\bibitem{Popov83} V. N. Popov, \textit{Functional Integrals in Quantum Field
Theory and Statistical Physics}, Part II (Pergamon, Oxford, 1980).

\bibitem{Beliaev} S. T. Beliaev, Sov. Phys. JETP \textbf{34}, 299 (1958).

\bibitem{Giamarchi99} T. Giamarchi and A. M. Tsvelik, Phys. Rev. B 59, 11398
(1999).

\bibitem{Jaime04} M. Jaime \textit{et al}, Phys. Rev. Lett. \textbf{93},
087203 (2004).

\bibitem{Suchitra05} S. E. Sebastian \textit{et al}, Phys. Rev. B \textbf{72} 
, 100404(R) (2005).

\bibitem{Ruegg06} Ch. R\"uegg \textit{et al.}, Phys. Rev. Lett. \textbf{98},
017202 (2007).

\bibitem{note0} A weak structural distortion observed at $\sim $ 100 K \cite 
{Samulon06} provides such a mechanism. Inelastic neutron scattering (INS)
indicates that the dimer axes is tilted, leading to a $10\% $ change of  the
intra-dimer exchange coupling $J$ (linear in the distortion) \cite{Ruegg06}.
The resulting non--frustrating interaction  $\sim 0.01J^{\perp }\simeq 10$
mK is quadratic in this distortion. We used preliminary INS measurements
that indicate $J^{\perp }\sim $ 1 K [Ch. Ruegg, private communication].

\bibitem{Prokofev02} N. Prokof'ev and B. Svistunov, Phys. Rev. A \textbf{66}
, 043608 (2002).

\bibitem{Rosch} O. R\"{o}sch, and M. Vojta, preprint, cond-mat/0702157.

\bibitem{Suchitra006} S. E. Sebastian \textit{et al}, Phys. Rev. B \textbf{74 
}, 180401(R) (2006).
\end{thebibliography}
\end{document}